# Dc SQUID based on a three-band superconductor with broken time-reversal symmetry

Y.S. Yerin[1], A.N. Omelyanchouk[1] and E. Il'ichev[2]

[1]B.Verkin Institute for Low Temperature Physics and Engineering
of the National Academy of Sciences of Ukraine
47 Lenin Ave., 61103 Kharkov, Ukraine

[2]Leibniz Institute of Photonic Technology, D-07702 Jena,
Germany

The behavior of a dc SQUID, based on a dirty point contacts between a single-band and three-band superconductor with broken time-reversal symmetry is investigated. Using earlier obtained results for Josephson effects in such systems new features in characteristics of a dc SQUID are revealed. It is shown that in the case of a BTRS (broken time-reversal symmetry) three-band superconductor for the applied external magnetic flux, which is divisible by the half-integer flux, strong degeneracy of ground states of a dc SQUID is taken place. This can lead to the appearance of possible multi-hysteresis loops on a dependence of a total flux in the dc SQUID from the externally applied flux. The number of these loops depends on the position of ground states of a three-band superconductor. Also it is found that dependencies of a critical current on applied magnetic flux can have complicated multi-periodic forms, which are differ from strictly periodic characteristics for conventional dc SQUIDs and Fraunhofer patterns for Josephson contacts in the external magnetic field.

## I. INTRODUCTION

Phase-sensitive studies in unconventional superconductors provide the most valuable information about the pairing symmetry of the order parameter. Among these experimental approaches a very fruitful method is the combination of a testing superconductor with a conventional s-wave isotropic superconductor into a superconducting quantum interference device (SQUID). This technique, also known as a SQUID interferometry, has proved itself already in the identification of the d-wave pairing symmetry in cuprate superconductors (see [1] and references therein]. So it is reasonable to expect the same succeeding implementation to recently discovered iron-based superconductors.

For these compounds since the discovery of the structure of the order parameter symmetry evolved from the so-called sign-reversal $s_\pm$-wave to the exotic chiral forms like s+id [2] and $s_\pm+is_{++}$ [3], still remaining debatable in present publications. In the last case the chirality of the order parameter can produce plethora of very interesting phenomena and related topological objects: collective modes, fractional vortices and states that broke time-reversal symmetry [4-16].

The broken time-reversal symmetry (BTRS) in the superconducting systems emerges when the phases of the order parameter cannot simultaneously satisfy to the condition of the energy minimum and thereby undergo frustration, creating several degenerated ground states of the superconductor. It should be note that at this moment a detection of the BTRS states in iron-based and other unconventional superconductors with a chiral structure of the order parameter is one of the most topical problems in the investigation of these compounds. Moreover, different tools have been already proposed for the detection of this phenomenon [15, 16].

Motivated by these unresolved challenges we address the problem of the dynamics of the dc SQUID with Josephson point contacts between a conventional s-wave and an isotropic dirty three-band s-wave superconductors. Using earlier obtained results on the Josephson effect in such superconducting systems [17], we investigate dependencies of the total flux in the dc SQUID from the externally applied flux (S states) and the magnetic field dependence of the critical current. We show, what new kind of features and effects for the characteristics of the dc SQUID we can expect due to the multi-band structure of the order parameter with possible manifestation of the BTRS state. Bearing in mind above formulated problems about the order parameter symmetry with potential realization of the BTRS, our consideration can be useful for the experimental investigation of these issues in iron-based and other multi-band unconventional superconductors, which demonstrate a multi-band superconducting state.

## II. MAIN EQUATIONS

We consider a dc SQUID with Josephson point contacts, which connect a s-wave single-band superconductor with a three-band one (fig. 1).

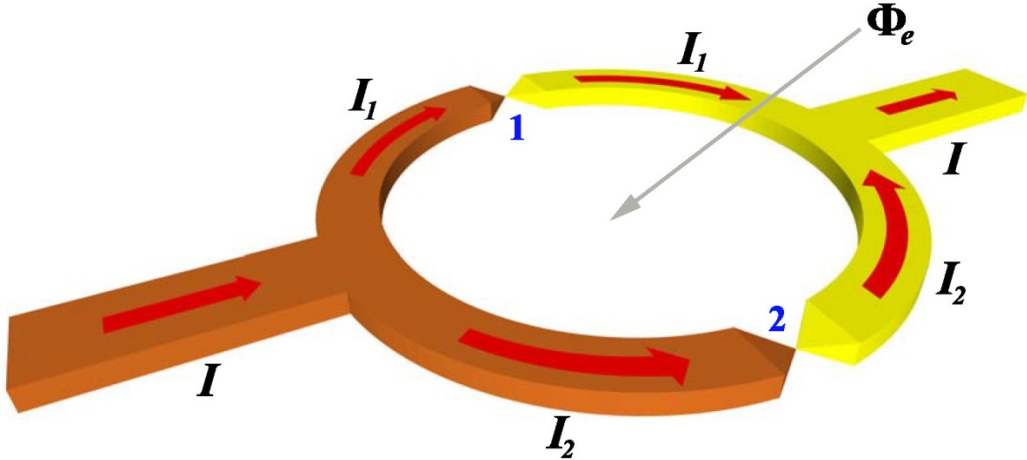

FIG. 1. (Color online) Schematic model of a dc SQUID based on point contacts (blue numbers 1 and 2) between a single-band (yellow part of the picture) and a three-band superconductor (brown part) with a bias current $I$ and applied magnetic flux $\Phi_e$. $I_1$ and $I_2$ both define appropriate currents through the point contacts 1 and 2.

It should be pointed out that a three-band superconductor can undergo a BTRS phenomenon and due to this reason can have two possible current-phase relations for the Josephson point contact with a conventional superconductor [17]. In general case these current-phase relations have cumbersome expressions (see Eq. (23) in [18] for the ballistic regime and Eq. (17) in [17] for the dirty limit).

Following the notations introduced earlier about phases of contacting superconductors (see Ref. 17) we denote the phase difference on the contact between the first order parameter of a three-band superconductor and the order parameter of the single-band superconducting part of the SQUID loop as $\chi_i$, where $i=1,2$ are index of the contact. The single-valuedness of the phase differences around the loop, which are thicker than the London penetration depth for the single- and three-band superconductors, requires

$$\chi_1 - \chi_2 = 2\pi \frac{\Phi}{\Phi_0}, \qquad (1)$$

where $\Phi$ is the total magnetic flux through the system and $\Phi_0$ is the flux quantum.

The same relation will be took place for the phase differences between the second, the third order parameter of the three-band superconductor and the order parameter of the single band one

$$(\chi_1 + \phi) - (\chi_2 + \phi) = 2\pi \frac{\Phi}{\Phi_0}, \qquad (2)$$

$$(\chi_1 + \theta) - (\chi_2 + \theta) = 2\pi \frac{\Phi}{\Phi_0}, \qquad (3)$$

where $\phi$ is the phase difference between the second and the first order parameters of a three band superconductor, and $\theta$ is the same for the third and the first ones. Phase differences $\phi$ and $\theta$ define the ground states of a three-band superconductor. If $\phi$ and $\theta$ are not equal to 0 or $\pi$ then the phase differences undergo a frustration and the BTRS state is realized in the three-band superconductor.

To obtain a full system of equations which describes the behavior of a dc SQUID, we need to supplement Eqs. (1)-(3) by relations for the bias current $I$ and full magnetic flux $\Phi$:

$$I = I_1 + I_2, \qquad (4)$$

$$\Phi = \Phi_e + L_1 I_1 - L_2 I_2, \qquad (5)$$

Here $I_i$ are currents, flowing through the contacts, $\Phi_e$ is the external magnetic flux, $L_i$ are inductances of the each branch of a dc SQUID. These inductances can be represented as $L_1 = \alpha L$ and $L_2 = (1-\alpha)L$, where $L$ is the full inductance of the loop [19].

Solving system of Eqs. (4) and (5) for $I_1$ and $I_2$ and taking into account quantization relations (1)-(3) we can rewrite Eqs. (4) and (5) in the dimensionless form:

$$i_1 = (1-\alpha)i + \frac{1}{\beta_{L1}}\left((\chi_1 - \chi_2) - \chi_e\right), \qquad (6)$$

$$i_2 = \alpha i - \frac{1}{\beta_{L1}}\left((\chi_1 - \chi_2) - \chi_e\right), \qquad (7)$$

where currents $i$, $i_1$ and $i_2$ are expressed in the units of the critical current of the first band of a three-band superconductor without interband interactions for the first point contact $I_{c1}^{(1)}$, $\beta_{L1} = \frac{2\pi L I_{c1}^{(1)}}{\Phi_0}$, the external flux $\chi_e = \frac{2\pi \Phi_e}{\Phi_0}$.

For simplicity we will investigate dirty superconductors at temperatures close to zero ($T=0$) and with almost identical to each other energy gaps, which, in turn, equal to $|\Delta|$. In this case currents $i_1$ and $i_2$ can be expressed as [17]:

$$i_1(\chi_1) = \cos\frac{\chi_1}{2}\operatorname{arctanh}\sin\frac{\chi_1}{2} + \frac{R_{N1}^{(1)}}{R_{N2}^{(1)}}\cos\frac{\chi_1+\phi}{2}\operatorname{arctanh}\sin\frac{\chi_1+\phi}{2} + \frac{R_{N1}^{(1)}}{R_{N3}^{(1)}}\cos\frac{\chi_1+\theta}{2}\operatorname{arctanh}\sin\frac{\chi_1+\theta}{2},$$

$$i_2(\chi_2) = \frac{R_{N1}^{(1)}}{R_{N1}^{(2)}}\cos\frac{\chi_2}{2}\operatorname{arctanh}\sin\frac{\chi_2}{2} + \frac{R_{N1}^{(1)}}{R_{N2}^{(2)}}\cos\frac{\chi_2+\phi}{2}\operatorname{arctanh}\sin\frac{\chi_2+\phi}{2} + \frac{R_{N1}^{(1)}}{R_{N3}^{(2)}}\cos\frac{\chi_2+\theta}{2}\operatorname{arctanh}\sin\frac{\chi_2+\theta}{2},$$

$$(8)$$

where $R_{N1}^{(i)}$, $R_{N2}^{(i)}$ and $R_{N3}^{(i)}$ are partial contributions from the each band to the normal resistance of the $i$-th contact. For non-equal gaps we will have cumbersome expressions for current-phase relations but on the qualitative level their structures do not differ from the case of coinciding energy gaps. So without loss of generality we can concentrate our attention on the above mentioned assumption.

Equations (6) and (7) can be derived from the variation of the energy $E$

$$E(\chi_1,\chi_2) = \frac{1}{2\beta_{L1}}\left((\chi_2-\chi_1)+\chi_e\right)^2 - i\left[(1-\alpha)\chi_1+\alpha\chi_2\right] + E_J(\chi_1,\chi_2), \quad (9)$$

as a function of $\chi_1$ and $\chi_2$. Here $E_J(\chi_1,\chi_2)$ is the Josephson energy of point contacts of a dc SQUID, which also depends on the temperature and the charge transfer regime (dirty or ballistic).

For the dirty limit and for $T=0$ we have

$$E_J(\chi_1,\chi_2) = \left(2\sin\frac{\chi_1}{2}\operatorname{arctanh}\sin\frac{\chi_1}{2}+\ln\cos^2\frac{\chi_1}{2}\right) + \frac{R_{N1}^{(1)}}{R_{N2}^{(1)}}\left(2\sin\frac{\chi_1+\phi}{2}\operatorname{arctanh}\sin\frac{\chi_1+\phi}{2}+\ln\cos^2\frac{\chi_1+\phi}{2}\right)$$

$$+ \frac{R_{N1}^{(1)}}{R_{N3}^{(1)}}\left(2\sin\frac{\chi_1+\theta}{2}\operatorname{arctanh}\sin\frac{\chi_1+\theta}{2}+\ln\cos^2\frac{\chi_1+\theta}{2}\right) + \frac{R_{N1}^{(1)}}{R_{N1}^{(2)}}\left(2\sin\frac{\chi_2}{2}\operatorname{arctanh}\sin\frac{\chi_2}{2}+\ln\cos^2\frac{\chi_2}{2}\right)$$

$$+ \frac{R_{N1}^{(1)}}{R_{N2}^{(2)}}\left(2\sin\frac{\chi_2+\phi}{2}\operatorname{arctanh}\sin\frac{\chi_2+\phi}{2}+\ln\cos^2\frac{\chi_2+\phi}{2}\right) + \frac{R_{N1}^{(1)}}{R_{N3}^{(2)}}\left(2\sin\frac{\chi_2+\theta}{2}\operatorname{arctanh}\sin\frac{\chi_2+\theta}{2}+\ln\cos^2\frac{\chi_2+\theta}{2}\right),$$

(10)

### III. BEHAVIOR OF A DC SQUID

By making use of the findings described above, below we will consider and describe the behavior of a dc SQUID (fig. 1) with identical point contacts (for simplicity), i.e. $\frac{R_{N1}^{(1)}}{R_{N2}^{(1)}} = \frac{R_{N1}^{(1)}}{R_{N3}^{(1)}} = \frac{R_{N1}^{(1)}}{R_{N1}^{(2)}} = \frac{R_{N1}^{(1)}}{R_{N2}^{(2)}} = \frac{R_{N1}^{(1)}}{R_{N3}^{(2)}} = 1$.

First of all we reconstruct the energy of a dc SQUID. Figure 2 presents contour plots of the energy as a function of the phase differences across the contacts $\chi_1$ and $\chi_2$ for a conventional dc SQUID (fig. 2 a and b) and a dc SQUID based on a three-band superconductor with the BTRS (fig. 2c-f) and non-BTRS state (fig. 2g-j) for zero magnetic flux (left column) and half-integer flux (right column).

In the case of a BTRS three-band superconductor for zero magnetic flux we observe only the shift of the position of an energy minimum of a dc SQUID from zero point $\chi_1 = \chi_2 = 0$ (like in a conventional dc SQUID) despite of the presence of the frustrated ground states (fig. 2c and 2e). The most remarkable features of a dc SQUID with a BTRS three-band superconductor are the strong degeneracy of the energy minimum (fig. 2d and fig. 2f) with the applying of the half-integer magnetic flux in comparison with a conventional dc SQUID (fig. 2b). With the increasing of the value $\beta_{L1}$ to the critical value $\beta_{L1}^{(c)}$ this degeneracy is removed, and the number of minima is decreased by a factor of two. According to our numerical simulations the value of $\beta_{L1}^{(c)}$ depends on the values of phases $\phi$ and $\theta$ in the ground state of a three-band superconductor. For instance, for the frustrated ground states with $\phi = 0.6\pi, \theta = 1.2\pi$ and $\phi = 1.4\pi, \theta = 0.8\pi$ the parameter $\beta_{L1}^{(c)}$ is found to be $\beta_{L1}^{(c)} \approx 13.04$.

For the case of a dc SQUID with a non-BTRS three-band superconductor we observe a contrary behavior. For zero magnetic flux $\Phi_e$ the degeneracy of an energy minimum is occurred (fig. 2g and fig 2i), despite the absence of the BTRS. In some way a similar behavior is taken place for the π-SQUID based on the weak links between a conventional s-wave superconductor and orthogonally oriented crystallographic planes of a d-wave one [20] however without appearing of the spontaneous magnetic flux. With the applying of the magnetic field the dynamics of the system has no features SQUID (fig. 1h and fig 1j) and a dc SQUID behave itself like a conventional SQUID.

We don't consider the energy of a dc SQUID based on a non-BTRS three-band superconductor with $\phi = \theta = 0$ because there are no qualitative changes in comparison with a conventional dc SQUID. Moreover below we will investigate the effect of a non-BTRS state on the characteristics of a dc SQUID only for a three-band superconductor with $\phi = 0, \theta = \pi$, $\phi = \pi, \theta = 0$ and $\phi = \pi, \theta = \pi$. We exclude from the consideration the ground state $\phi = 0, \theta = 0$ due to the possibility to reduce this case to a single-band superconductor with the triple value of the energy gap.

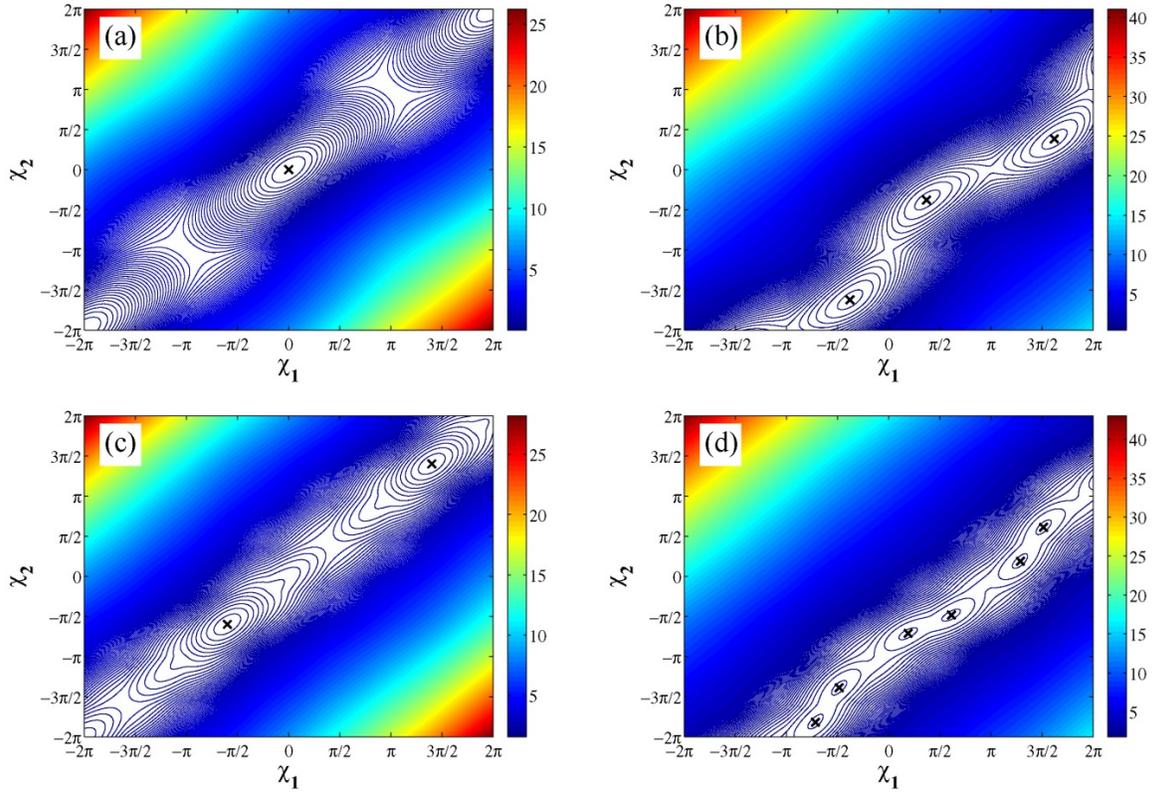

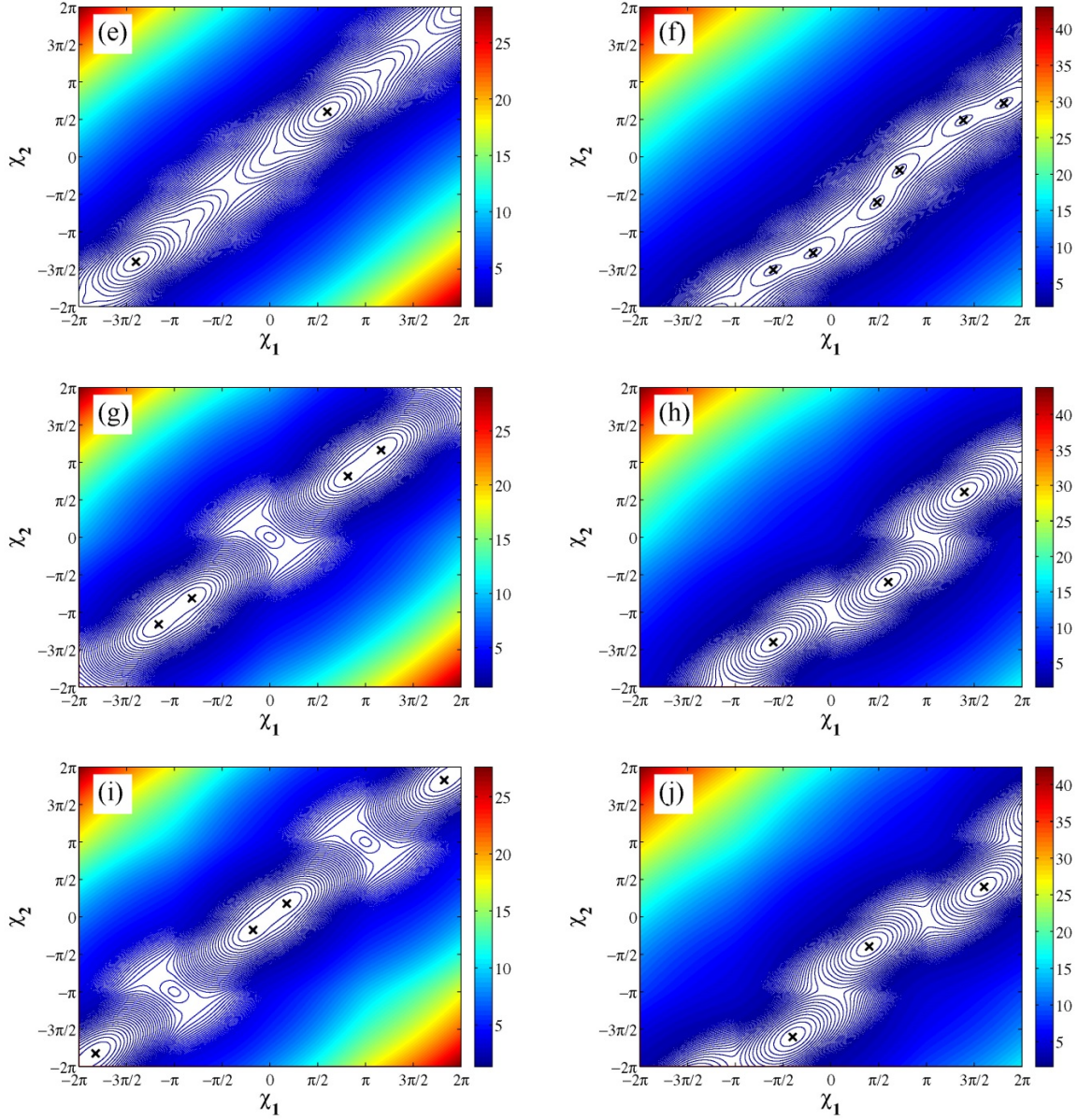

FIG. 2. (Color online) Contour plots of the energy of the dc SQUID for zero external magnetic flux $\chi_e = 0$ ($\Phi/\Phi_0 = 0$, left column) and for $\chi_e = \pi$ ($\Phi/\Phi_0 = 0.5$, right column) in the absence of a bias current. (**a**) and (**b**) are plotted for the point contacts between s-wave single-band superconductors; (**c**), (**d**) and (**e**), (**f**) are plotted for the point contacts between a single-band and a BTRS three-band superconductors for the frustrated ground states $\phi = 0.6\pi, \theta = 1.2\pi$ and $\phi = 1.4\pi, \theta = 0.8\pi$ correspondingly; (**g**), (**h**) and (**i**), (**j**) are the same for a non-BTRS superconductor with the ground state $\phi = \pi, \theta = \pi$ or $\phi = 0, \theta = \pi$ correspondingly. Crosses indicate positions of the minimum. For all dc SQUIDs $\beta_{L1} = 3$.

We start the consideration of characteristics and properties of a dc SQUID from the case with a vanishing inductance of the loop, when the full magnetic flux equals to the external flux

$$\chi_1 - \chi_2 = 2\pi \frac{\Phi_e}{\Phi_0}. \tag{11}$$

The problem of the finding of the critical current $i_c$ for a dc SQUID versus the applied magnetic flux $i_c = i_c\left(\frac{\Phi_e}{\Phi_0}\right)$ is reduced to the determination of the maximum value of the function of the total current

$$i(\chi_1, \chi_2) = i_1(\chi_1) + i_2(\chi_2), \qquad (12)$$

where we have to take into account the quantization condition (11). Here $i_1(\chi_1)$ and $i_2(\chi_2)$ are determined by Eqs. (8). Numerical solution of this problem is represented in Fig. 3 for the identical point contacts in the case of a BTRS three-band superconductor (fig. 3a) and a non-BTRS one (fig. 3b).

We found that in the presence of the BTRS, despite of two possible different current-phase relations [17], dependencies $i_c = i_c\left(\frac{\Phi_e}{\Phi_0}\right)$ will be the same for the both ground states of a three-band superconductor (Fig. 3a). The same situation is realized for a non-BTRS three-band superconductor with the ground states $\phi = 0, \theta = \pi$ and $\phi = \pi, \theta = \pi$ (fig. 3b).

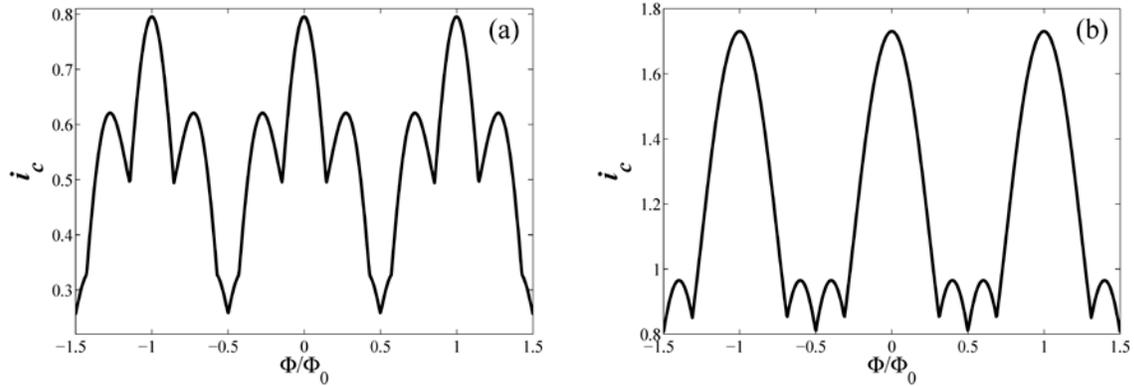

FIG. 3. Dependence of a critical current of the symmetrical dc SQUID with a vanishing inductance on the applied magnetic flux for a BTRS three-band superconductor with the frustrated ground states $\phi = 0.6\pi, \theta = 1.2\pi$ and $\phi = 1.4\pi, \theta = 0.8\pi$ (**a**) and a non-BTRS three-band superconductor with $\phi = 0, \theta = \pi$ and $\phi = \pi, \theta = \pi$ (**b**).

Comparing figure 3a and figure 3b we can conclude that in the case of a BTRS state the critical current dependence of a dc SQUID has more pronounced side peaks than for a non-BTRS superconductor.

Incorporation of the asymmetry in the critical currents of point contacts leads to the asymmetry for the dependence of the critical current on the external magnetic flux for the BTRS and non-BTRS three-band superconductor (Fig. 4), increasing the difference in the form of these curves. From the experimental point of view it means that the critical current of the system can depend on the direction of the applied magnetic flux.

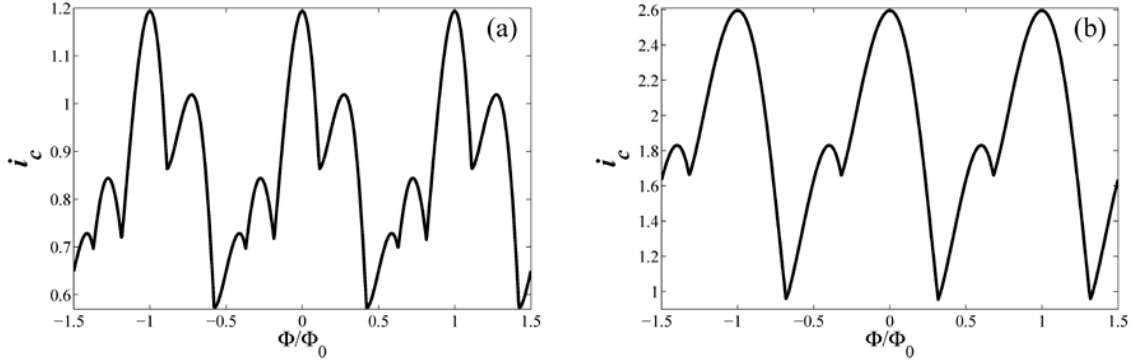

FIG. 4. Critical current of the asymmetrical ( $\frac{R_{N1}^{(1)}}{R_{N2}^{(1)}} = \frac{R_{N1}^{(1)}}{R_{N3}^{(1)}} = 1$ and $\frac{R_{N1}^{(1)}}{R_{N1}^{(2)}} = \frac{R_{N1}^{(1)}}{R_{N2}^{(2)}} = \frac{R_{N1}^{(1)}}{R_{N3}^{(2)}} = 2$ ) dc SQUID with a vanishing inductance on the applied magnetic flux for a BTRS three band superconductor with frustrated ground states $\phi = 0.6\pi, \theta = 1.2\pi$ and $\phi = 1.4\pi, \theta = 0.8\pi$ (**a**) and non-BTRS three-band superconductor with $\phi = 0, \theta = \pi$ and $\phi = \pi, \theta = \pi$ (**b**).

In the case of iron-based superconductors complicated behavior of the critical Josephson current vs. applied magnetic flux has been already observed for the edge-type Josephson junctions in Ba-122 thin films [21, 22]. So, it would be quite interesting to check this asymmetry for a dc SQUID based on 122 family of iron-based superconductors, which are considered as the most promising candidate for the detecting of a BTRS symmetry phenomenon.

Now we proceed to the consideration of the S states (dependencies of the total flux in the dc SQUID from the externally applied flux) for zero bias current $i = 0$. In the case of identical point contacts we can reduce a dc SQUID to an equivalent rf SQUID with the phase difference over the single contact $\chi_{rf}$, which is connected with $\chi_1$ and $\chi_2$ by the relations

$$\chi_1 = \Delta\chi + \chi_{rf}, \tag{13}$$
$$\chi_2 = \Delta\chi - \chi_{rf}, \tag{14}$$

The parameter $\Delta\chi$ can be found after summation of (6) and (7)

$$\cos\frac{\chi_1}{2}\operatorname{arctanh}\sin\frac{\chi_1}{2} + \cos\frac{\chi_1+\phi}{2}\operatorname{arctanh}\sin\frac{\chi_1+\phi}{2} + \cos\frac{\chi_1+\theta}{2}\operatorname{arctanh}\sin\frac{\chi_1+\theta}{2} + \\ \cos\frac{\chi_2}{2}\operatorname{arctanh}\sin\frac{\chi_2}{2} + \cos\frac{\chi_2+\phi}{2}\operatorname{arctanh}\sin\frac{\chi_2+\phi}{2} + \cos\frac{\chi_2+\theta}{2}\operatorname{arctanh}\sin\frac{\chi_2+\theta}{2} = 0. \tag{15}$$

For conventional dc SQUID the solution of Eq. (15) is satisfied, when the parameter $\Delta\chi$ equals to 0 or π. But in general case $\Delta\chi$ depends on values of $\phi$ and $\theta$ and can be found numerically only. For a non-BTRS three-band superconductor Eq. (15) has two solutions, which coincide with the solutions for a conventional dc SQUID, i.e. $\Delta\chi = 0$ and $\Delta\chi = \pi$.

After change of phase variables $\chi_1$ and $\chi_2$ for $\chi_{rf}$ Eqs. (6) and (7) are transformed to the single equation

$$\chi_{rf} + \frac{1}{2}\beta_{L1}\left(\cos\frac{\chi_{rf}+\Delta\chi}{2}\operatorname{arctanh}\sin\frac{\chi_{rf}+\Delta\chi}{2} + \cos\frac{\chi_{rf}+\Delta\chi+\phi}{2}\operatorname{arctanh}\sin\frac{\chi_{rf}+\Delta\chi+\phi}{2} \right.$$
$$\left. +\cos\frac{\chi_{rf}+\Delta\chi+\theta}{2}\operatorname{arctanh}\sin\frac{\chi_{rf}+\Delta\chi+\theta}{2}\right) = \frac{1}{2}\chi_e \quad (16)$$

Solving (16) we find the S-states of a dc-SQUID for a BTRS (Fig. 5b) and a non-BTRS (Fig. 6b) three-band superconductor and compare them with a conventional dc-SQUID with a dirty point contact also at $T = 0$ (Fig. 5a and 5b). We should note that for a conventional dc SQUID there is the second stable solution (dashed blue lines in Fig. 5a and 5b) due to two solutions of Eq. (15). For a dc SQUID with a BTRS three-band superconductor another kind of the stable S-states is also existed. For a given ground state of a three-band superconductor these stable S-states correspond to the S-states of the second ground state and can be obtained from the solution of Eq. (16) by substitution of $\Delta\chi \to \Delta\chi + \pi$.

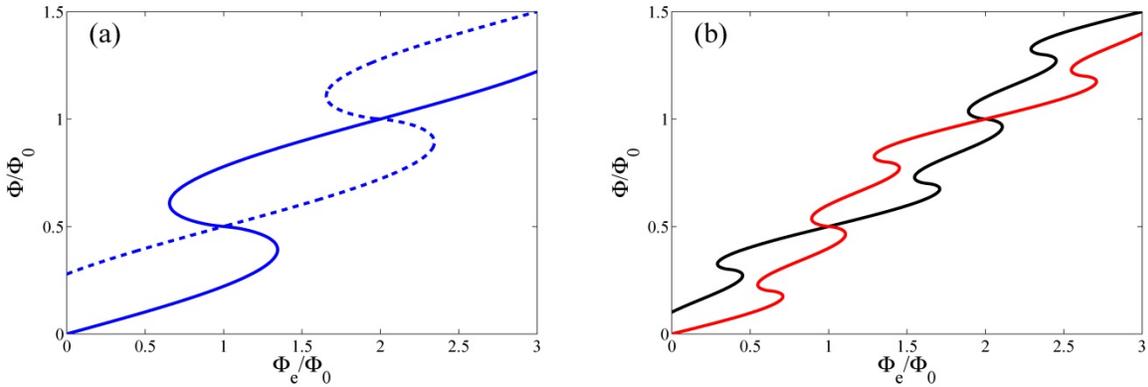

FIG. 5. (Color online) S-states in a conventional dc SQUID (**a**) and in a dc SQUID with a BTRS three-band superconductor (**b**). Blue solid and dashed lines demonstrate two kind of possible S-states of a conventional dc-SQUID. S-states in the case of a BTRS three-band superconductor correspond to $\phi = 0.6\pi, \theta = 1.2\pi$ with $\Delta\chi \approx 1.25664 \pm 2\pi n$ (black line) and $\phi = 1.4\pi, \theta = 0.8\pi$ with $\Delta\chi \approx \pi \pm 2\pi n - 1.25664$ (red line), $n \in \mathbb{Z}$. For all curves $\beta_{L1} = 3$.

In the case of a non-BTRS three-band superconductor with the ground state $\phi = 0, \theta = \pi$ ($\phi = \pi, \theta = \pi$) we can obtain the second stable solution for the S-states of a dc SQUID again after substitution of $\Delta\chi \to \Delta\chi + \pi$ in Eq. (16). These solutions coincide with the S-states for $\phi = \pi, \theta = \pi$ ($\phi = 0, \theta = \pi$).

According to our analysis as in the case of the conventional dc SQUID the S-states will be stable only if the first derivative $d\Phi/d\Phi_e$ is positive, in other words the stable states are the S-states with a positive slope of dependencies $\Phi = \Phi(\Phi_e)$.

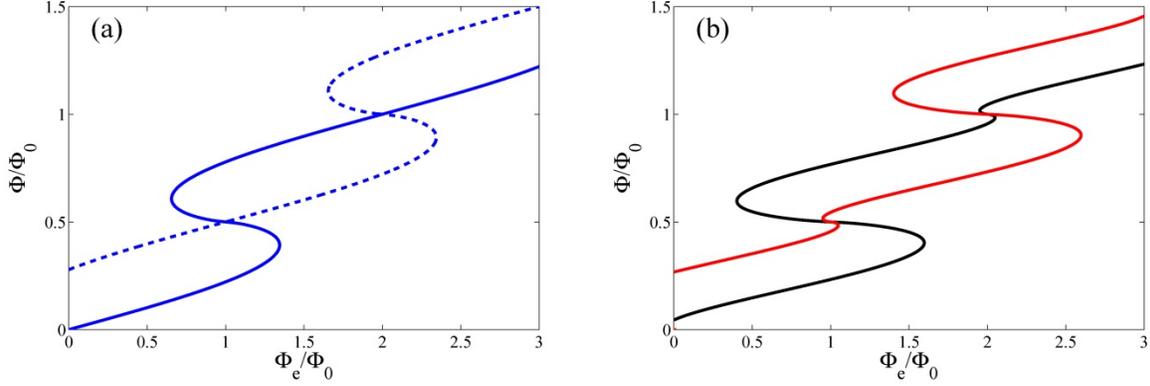

FIG. 6. (Color online) S-states in a conventional dc SQUID (**a**) and in a dc SQUID with a non-BTRS three-band superconductor (**b**). Blue solid and dashed lines demonstrate two kind of possible S-states of a conventional dc-SQUID. For the case of a non-BTRS three-band superconductor the S-states corresponds to $\phi = 0, \theta = \pi$ with $\Delta\chi = 0$ (black line) and for $\phi = \pi, \theta = \pi$ with $\Delta\chi = 0$ (red line). For all curves $\beta_{L1} = 3$.

In comparison with a hysteretic behavior of a conventional dc SQUID we have complicated multi-hysteretic S-states of a dc SQUID with a BTRS three-band superconductor. In other words during measurements of full magnetic flux vs. applied one, we can observe additional jumps on these dependencies. The latter can be considered as the feature of a dc SQUID with a BTRS three-band superconductor.

## IV. CONCLUSIONS

To summarize we have investigated theoretically the dynamics of a dc SQUID based on point contacts between a conventional single-band and three-band superconductor with and without BTRS. We have revealed new features in the behavior of a dc SQUID and have found qualitative and quantitative difference in the characteristics of such device, based on a BTRS and non-BTRS three-band superconductor. We have shown that dependencies of a critical current and total magnetic flux versus applied magnetic flux in a dc SQUID can be considered as experimental tools for the detection of BTRS states in multi-band superconductors.

### Acknowledgments

This work was supported by DKNII (Project No. M/231-2013) and by BMBF (UKR-2012-028).

### REFERENCES

1. C. C. Tsuei and J. R. Kirtley, Rev. Mod. Phys. 72, 969 (2000).

2. Wei-Cheng Lee, Shou-Cheng Zhang, and Congjun Wu, Phys. Rev. Lett. 102, 217002 (2009).

3. V. Stanev and A.E. Koshelev, Phys. Rev. B 89, 100505(R) (2014).

4. D. F. Agterberg, V. Barzykin, and L. P. Gor'kov, Phys. Rev. B 60, 14868 (1999).

5. V. Stanev and Z. Tesanovic, Phys. Rev. B 81, 134522 (2010).

6. Y. Tanaka and T. Yanagisawa, J. Phys. Soc. Japan 79, 114706 (2010).

7. J. Carlstrom, J. Garaud, and E. Babaev, Phys. Rev. B 84, 134518 (2011).

8. R. G. Dias and A. M. Marques, Supercond. Sci. Technol. 24, 085009 (2011).

9. T. Yanagisawa, Y. Takana, I. Hase, and K. Yamaji, J. Phys. Soc. Jpn. 81, 024712 (2012).

10. X. Hu and Z. Wang Phys. Rev. B 85, 064516 (2012).

11. Shi-Zeng Lin and Xiao Hu, New J. Phys. 14, 063021 (2012).

12. S. Z. Lin and X. Hu, Phys. Rev. Lett. 108, 177005 (2012).

13. T. Bojesen, E. Babaev, A. Sudbø, Phys. Rev. B 88, 220511(R) (2013)

14. S. Maiti and A. V. Chubukov, Phys. Rev. B 87, 144511 (2013).

15. M. Marciani, L. Fanfarillo, C. Castellani, and L. Benfatto, Phys. Rev. B 88, 214508 (2013).

16. J. Garaud and E. Babaev, Phys. Rev. Lett. 112, 017003 (2014).

17. Y.S. Yerin, A.N. Omelyanchouk, Low Temp. Phys. 40, 943 (2014).

18. Z. Huang and X. Hu, Appl. Phys. Lett. 104, 162602 (2014).

19. J. Clarke, W. M. Goubau, and M. B. Ketchen, J. Low. Temp. Phys. 25, 99 (1976).

20. H. J. H. Smilde, Ariando, D. H. A. Blank, H. Hilgenkamp, and H. Rogalla, Phys. Rev. B 70, 024519 (2004).

21. S. Schmidt, S. Döring, F. Schmidl, V. Grosse, P. Seidel, K. Iida, F. Kurth, S. Haindl, I. Mönch, B. Holzapfel, Appl. Phys. Lett. 97, 172504 (2010).

22. S. Döring, S. Schmidt, F. Schmid, V. Tympe, S. Haind, F. Kurth, K. Iida, I. Mönch, B. Holzapfel and P. Seidel, Supercond. Sci. Technol. 25 084020 (2012).